\documentclass[aps,prapplied,reprint,floatfix,longbibliography]{revtex4-1}

\usepackage{amsmath,amsfonts,amssymb,graphicx,braket,qcircuit} 
\usepackage[]{units} 
\usepackage{hyperref} \hypersetup{ colorlinks=true, linkcolor=blue, filecolor=blue, urlcolor=blue, citecolor=blue }

\begin{document} 
\title{Hybrid exchange--measurement-based qubit operations in semiconductor double-quantum-dot qubits} 
\author{Matthew Brooks} 
\email{matthew.brooks@lps.umd.edu} 
\author{Charles Tahan} \affiliation{Laboratory for Physical Sciences, 8050 Greenmead Dr., College Park, MD 20740, USA}

%\date{\today}
\begin{abstract}

Measurement-based quantum computing (MBQC) promises an alternative approach to quantum computation that has natural compatibility with error correction codes at the cost of a polynomial increase in physical qubits. MBQC implementations have previously focused on photonic systems where two-qubit gates are difficult. On the other hand, semiconductor spin qubit systems offer fast two-qubit gates via the exchange interaction. To explore the benefits of MBQC on spin systems, two hybrid measurement-exchange schemes for full qubit control with semiconductor double quantum dot spin qubits are considered. Protocol 1 fully realizes the cluster-state approach to MBQC but requires singlet-triplet qubits in a magnetic field gradient. Protocol 2 implements a direct measurement based replacement for more traditional gate-based encoded operations, without need for magnetic field gradients. Both protocols demonstrate full single and two qubit control through a combination of inter- and intra- qubit exchange and measurement onto the singlet-triplet basis. We show that both schemes suppress individual qubit spin-state leakage errors and offer fast gate times, up to known phase and Pauli corrections. 
\end{abstract}
\maketitle

\section{Introduction} \label{sec:Intro}

Measurement-based quantum computing (MBQC) performs unitary rotations of encoded quantum states in a quantum processor by taking measurements, projective or partial, of qubits within entangled systems\cite{raussendorfQuantumComputingMeasurements2000,verstraeteValencebondStatesQuantum2004,raussendorfOnewayQuantumComputer2002,nielsenUniversalQuantumComputation2003,raussendorfMeasurementbasedQuantumComputation2003,jozsaIntroductionMeasurementBased2005,browneOnewayQuantumComputation2006}. The amount of calculations that can be performed on an MBQC processor is limited by the amount of entangled qubits in the initial state known as the resource state. This allows for parallelization of some gates on such a processor by simultaneously performing commuting measurements of qubits within the resource state\cite{raussendorfQuantumComputingMeasurements2000,verstraeteValencebondStatesQuantum2004,raussendorfOnewayQuantumComputer2002,nielsenUniversalQuantumComputation2003,raussendorfMeasurementbasedQuantumComputation2003,jozsaIntroductionMeasurementBased2005,browneOnewayQuantumComputation2006,raussendorf2007fault,van2006universal,plato2008random,shettell2020graph}. This is unlike more conventional quantum circuit-based models realized in solid state systems where unitaries are achieved by driving interactions between qubits\cite{barends2014superconducting,arute2019quantum,preskill2018quantum,wang201816}. In such systems like superconducting circuits\cite{arute2019quantum,wang201816}, trapped ions\cite{zhang2017observation,ballance2016high} or semiconductor quantum dots (QDs)\cite{hanson2007spins,russ2017three,mi2018coherent,veldhorst2017silicon}, qubit control typically stems from interactions with externally-driven electric, magnetic or laser fields. 

Semiconductor spin qubits offer long coherence times and precise control of inter-QD energy detuning and tunnel couplings\cite{takeda2020resonantly,yang2019silicon,zajac2018resonantly,russ2018high} within both commercially available and isotopically purifiable semiconductors, making them excellent candidates for quantum processors. Exchange interactions allow for fast spin-spin entangling operations that are susceptible to errors due to charge noise in the controlling gates\cite{russ2017three,reed2016reduced,yoneda2018quantum}. Therefore simple encoded qubits of two or more spins\cite{levy2002universal,johnson2005triplet,laird2010coherent,russ2017three} are employed to protect against charge noise errors, at the cost of susceptibility to leakage errors. Leakage errors occur when the qubit couples to spin states outside the encoded qubit subspace and are generally caused by inter-qubit exchange interaction\cite{wardrop2014exchange} or hyperfine interactions\cite{andrews2019quantifying}, depending on the encoded subspace. In this work we are interested in exploring MBQC with encoded spin qubits to avoid these issues by demonstrating full qubit control, within schemes that suppress qubit leakage. Such implementations are inherently compatible with error correction schemes at the cost of a polynomial increase in the required qubits\cite{raussendorfMeasurementbasedQuantumComputation2003,raussendorfFaulttolerantOnewayQuantum2006}.

Typically, experimental realizations of MBQC are in photonic systems\cite{bell2014experimental,russo2018photonic,bodiya2006scalable}, although there has been some experimental success in trapped ion systems\cite{lanyonMeasurementbasedQuantumComputation2013}. The challenges of photonic-based QC (lack of a two qubit interaction) have traditionally motivated MBQC while the assumed large physical qubit resource overhead has discouraged exploration of solid-state MBQC processors. There has been some work combining photonics with QDs in cavities to build cluster states with spatially separated solid state qubits manipulated optically\cite{benjamin2009prospects}, however this is dependent on preparation by measuring the random decay of the cavities, and does not levy the advantage of fast spin qubit gates with exchange. Additionally, there has been recent work in MBQC by only fermionic singlet-triplet measurement\cite{freedman2021symmetry}, however this method lacks analogue gate control, is not implemented in an encoded subspace and cannot take advantage of exchange interactions. A notable exception in purely solid-state implementations of quantum processors are charge parity measurement-based anyonic braiding operations in Majorana fermion qubits coupled to QDs\cite{karzigScalableDesignsQuasiparticlePoisoningProtected2016,MeasurementBasedControlled,PhysRevLett,grimsmo2019majorana}. This scheme however is functionally equivalent to a typical gate-based approach as there is no larger resource state employed, thus there is less of a qubit overhead problem bar the ancilla QDs coupling the logical Majorana qubits.

Here we discuss two protocols of performing MBQC with semiconductor spin qubits assumed to be given by Si double-quantum-dot (DQD) qubits\cite{zwanenburg2013silicon,zajac2018resonantly}. In the first protocol, 1 and 2D cluster states in spin qubits are proposed. This scheme offers efficient arbitrary single qubit state preparation, as well as full single and two qubit control and compatibility with stabilizer codes by design of static qubit graph geometries. The second protocol employs entangling measurements to achieve a hybrid scheme more akin to conventional gate-based quantum computation, similar to the Majorana-qubit measurement gates schemes. Both schemes are robust against spin state leakage and offer fast gate times.

%(reth paragrpah, is this needed)
%Both of these protocols are unique to spin qubits and offer alternative paradigms to QC, beyond the gate-based quantum analog to classical information processing. The suppression of leakage and driving noise offers immediate improvements to the NISQ scale devices with promise of scaling to larger processors with by writing and reading states to and from simple stabliser codes.

This paper is organized as follows, firstly the ingredients of cluster state MBQC exploited by one of our proposals are introduced in Sec.~\ref{sec:MBQC}. Then, our first hybrid exchange-MBQC control scheme is given in Sec.~\ref{sec:Hybrid}, followed by the second protocol employing entangling measurements in Sec.~\ref{sec:Quantum Bus}. In Sec.~\ref{sec:CurveMeas} the methods of qubit measurement are discussed and in Sec.~\ref{sec:leak} the protection of the protocols to leakage is discussed. Then, lastly, the results are summarized in Sec.\ref{sec:Discussion}.
 
\section{Cluster States} \label{sec:MBQC}

The foundation of MBQC are quantum processors made up of large entangled states, such as a cluster state\cite{raussendorfMeasurementbasedQuantumComputation2003,raussendorfFaulttolerantOnewayQuantum2006}.  A cluster state $\ket{\Psi}$ is given by 
\begin{equation}
	\ket{\Psi}=\prod_{j\in N(i)}CZ\ket{+_i}\otimes\ket{+_j} 
\end{equation}

\noindent where $N(i)$ gives all adjacent interacting qubits to qubit $i$, $CZ$ gives the entangling two qubit control-phase ($CZ$) gate diag$\{1,1,1,-1\}$ and $\ket{\pm}=(\ket{0}\pm \ket{1})/\sqrt{2}$ are the eigenstates of the Pauli-$x$ ($\sigma_x$) matrix. In the simplest geometry, a 1D array of qubits each entangled by nearest neighbor, the first of which is in some prepared state $\ket{\psi}$, single-qubit rotations on $\ket{\psi}$ are achieved by measuring each qubit along the array sequentially in the $\ket{M_\pm(\theta)}=(\ket{0}\pm e^{-i\theta}\ket{1})/\sqrt{2}$ basis. Each measurement performs the following rotation 
\begin{equation}
	\ket{\tilde{\psi}}=\sigma_x^s H P(\theta)\ket{\psi}=\sigma_x^s W(\theta)\ket{\psi} 
\end{equation}

\noindent on the data qubit $\ket{\psi}$, teleporting the state onto the neighboring qubit,  where $s$ is the measurement outcome ($s=0,1$ for $\ket{M_\pm(\theta)}$), $H$ is the Hadamard gate and $P(\theta)$ is a phase gate, the phase of which depends on the angle of the rotated measurement basis. The $\sigma_x^s$ term is a Pauli correction dependent on the measurement outcome. To perform a controlled arbitrary single qubit rotation, a maximum of four sequential measurements accounting for the previous Pauli correction are needed since $H$ and $\sigma_x$ do not commute. Experimentally, performing an $\ket{M_\pm(\theta)}$ measurement is equivalent to a Pauli-$x$ measurement ($\ket{M_\pm(0)}$) between phase gates, i.e. $P(\theta)\ket{M_\pm(0)}\bra{M_\pm(0)}P(-\theta)$.

Using the same tools to achieve single qubit control, two qubit rotations can be done in 2D arrays of cluster states. For example, a control-not ($CX$) gate requires only a resource state of four qubits arranged by three in a line containing the target qubit and the fourth entangled to the middle qubit as the control qubit. By two simultaneous $\ket{M_\pm(0)}$ measurements on the wire, the target is teleported along the wire whilst the control qubit is stationary. Therefore if the necessary ingredients for full single qubit control can be given for a cluster state scheme, two qubit control can be assumed to be achievable if a 2D graph can be initialized.

\begin{figure}
	[!ht] 
	\includegraphics[width=\linewidth]{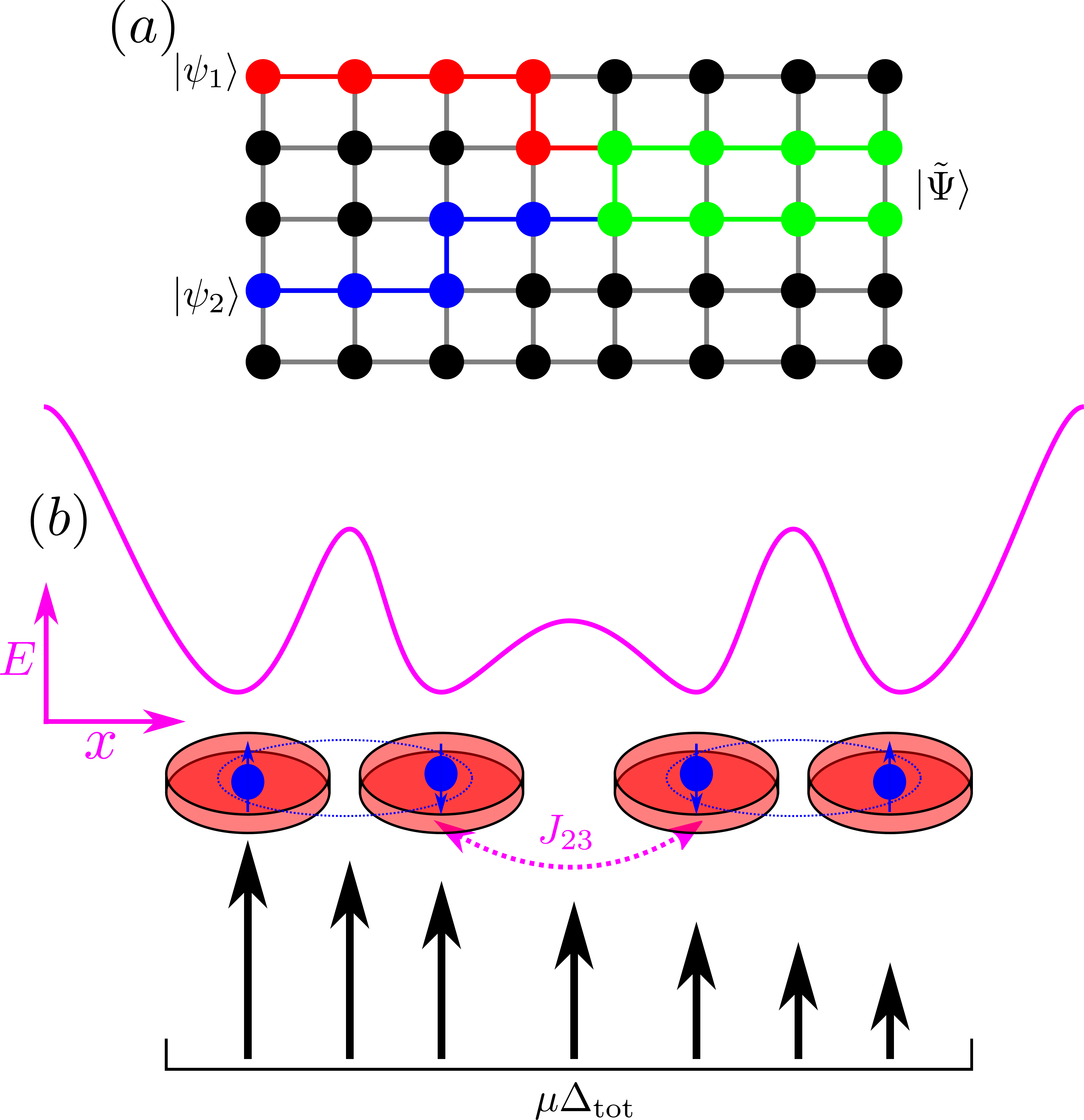} 
	\caption{(a) Grid cluster state, consisting of qubits (black dots) all initialized in the $\ket{+}$ state, entangled to their neighboring qubits by a $CZ$ drawn as a black line. Two logical qubit states $\ket{\psi_1}$ (red) and $\ket{\psi_2}$ (blue) start on the left hand side of the graph, and follow a path of measurements through the graph given by their respective color. The qubits are then entangled  by a two qubit gate measurement sequence outputting the state $\ket{\tilde{\Psi}}$, and all unused qubits in the graph can be disentangled from the output state by Pauli-$z$ measurements. (b) Linear two semiconductor double quantum dot qubits diagram. Each qubit consists of two dots (red) is charged with one electron (blue) in a $m_z$-0 singlet-triplet spin configuration (blue dotted line). The two adjacent dots from each qubit are coupled by en exchange interaction $J_{23}$ (magenta dotted line) given by the depicted potential shape (magenta line). The magnetic field gradients for each qubit as well as the whole system are given (black).} 
\label{fig:DQD} \end{figure}

\section{Hybrid Exchange-Measurement Graph Protocol} \label{sec:Hybrid}

\subsection{Qubit Model: Magnetic gradients} \label{sec:gradqub}

In the first protocol, the qubits considered are assumed to be Si DQD spin qubits charged with two electrons, one in each dot in a $m_z=0$ singlet-triplet $\ket{s}(\ket{t_0})=(\ket{\uparrow\downarrow}\pm\ket{\downarrow\uparrow})/\sqrt{2}$ configuration lifted from the $m_z=\pm1$ triplet states $\ket{t_+}(\ket{t_-})=\ket{\uparrow\uparrow}(\ket{\downarrow\downarrow})$ by an external magnetic $B_0$ field. Additionally, the qubits are assumed to experience static magnetic field gradients, implemented by micromagnets\cite{mcneil2010localized}, or engineered $g$-factors\cite{ruskov2018electron,harvey2019spin}, across the two dots. This allows for exchange-based $CZ$ gates to be implemented between two tunnel-coupled qubits in the limit of weak tunneling $t_{ij}\ll E_c$ (where $t_{ij}$ is the tunnel coupling between dots $i$ and $j$, $E_c$ is the charging energy of the dots) and the exchange-interaction $J_{ij}$ between the dots satisfies $J_{ij}\ll B_0$. In this limit the exchange interaction between dots is given by $J_{ij}=4t_{ij}^2/E_c$. This is described for two qubits coupled linearly by the following exchange effective Hamiltonian (assuming $\hbar=1$ throughout)\cite{wardrop2014exchange}
\begin{equation}
	H_{\text{eff}}=(\mu \Delta_{12}+\tilde{B})\sigma_z^1+(\mu \Delta_{34}+\tilde{B})\sigma_z^2-\frac{1}{4}J_{23}\left(\sigma_z^1 \sigma_z^2+\mathbb{I}\right) 
\label{eq:Heff} \end{equation}

\noindent where $\mu$ is the Bohr magneton, $\Delta_{ij}=B_i-B_j$ are the intra-qubit magnetic field gradients where $B_i$ is the magnetic field experienced by the $i^{\text{th}}$ dot, and $\tilde{B}$ is given as 
\begin{equation}
	\begin{split}
		\tilde{B}=\frac{1}{2}\left[-\frac{1}{2}\left(\mu \Delta_{12}+\mu \Delta_{34}\right)+\mu \Delta B-\right. \\
		\left. -\sqrt{\left[\mu \Delta_B-\frac{1}{2}\left(\mu \Delta_{12}+\mu \Delta_{34}\right)\right]^2+\frac{1}{4}J_{23}^2}\right] 
	\end{split}
	\label{eq:Btilde} \end{equation}

\noindent where $\Delta B$ is the inter-qubit magnetic field difference, i.e. $\Delta B=(B_1+B_2-B_3-B_4)/2$, and $\sigma_z^i$ is the Pauli-$z$ matrix of the $i^{\text{th}}$ qubit in the $x$-basis of a singlet-triplet qubit, i.e. $\sigma_z=\ket{0}\bra{0}-\ket{1}\bra{1}$ where $\ket{0}=\left(\ket{t_0}+\ket{s}\right)/\sqrt{2}=\ket{\uparrow\downarrow}$ and $\ket{1}=\left(\ket{t_0}-\ket{s}\right)/\sqrt{2}=\ket{\downarrow\uparrow}$. Assuming all QDs global magnetic field gradient such that $\Delta_B=\Delta_{23}$ and $\Delta_{12}=\Delta_{34}=\Delta$, assuming the qubits to be identical and $\Delta_{23}$ is given by the distance between adjacent qubits, a $CZ$ is achieved between the two qubits when
\begin{equation}
	\tau=\frac{\pi+4n_1}{4\mu\Delta+2\mu\Delta_{23}-\sqrt{J_{23}^2+(2\mu\Delta_{23})^2}} 
\end{equation}

\noindent when the inter-qubit exchange satisfies 
\begin{equation}
	\begin{split}
			&J_{23}=-\frac{\mu(4 n_1+4 n_2+1)}{4 n_2 (4 n_1+2 n_2+1)} \left[\Delta_{23}+\Delta  (8 n_1+2)\right.  \\& \left.+4 \Delta_{23} n_1+\sqrt{(\Delta_{23}+4 \Delta_{23} n_1)^2+4 \Delta ^2 (4 n_1+4 n_2+1)^2}\right.  \\& \left.\overline{+4 \Delta  \Delta_{23} (4 n_1+4 n_2+1)^2}\right]
	\end{split}
\end{equation}

\noindent where $n_1$ and $n_2\in \mathbb{Z}$. A two qubit DQD with this geometry is depicted in Fig.~\ref{fig:DQD}(b). These expressions have been written with the integer constants as they demonstrate how pulse times and strengths allow for leakage to be suppressed in a gate based regime as leakage of the qubits into spin states outside of the computational basis (i.e. $\ket{\uparrow\uparrow}$ and $\ket{\downarrow\downarrow}$) is suppressed if $J_{23}\ll\mu\Delta$\cite{wardrop2014exchange}. As will be evident in Sec.~\ref{sec:leak}, this condition is not needed to suppress leakage in the following proposal. Alternatively, an exchange based $CZ$ gate can be achieved by time-crystal-inspired alternating inter- and intra-qubit exchange pulses\cite{van2021protecting}, however this will not be considered in this work.

It can be assumed that the qubits in both proposed protocols are operated at their symmetric operating point (SOP)\cite{reed2016reduced}. The SOP is the point at which $\partial E_{s/t}/\partial \epsilon =0$, where $E_{s/t}$ are the eigenvalues of the singlet/triplet spin-states and $\epsilon$ is the detuning between the two dots of the qubit. Operating here protects against first-order charge noise.

Armed with a $CZ$ gate, the necessary ingredients for initializing a cluster state are given. Firstly, the qubits are initialized into a spin-singlet state (a $\ket{-}$ in the qubit space) by charging two electrons into one dot of each qubit and tuning the intra-qubit tunnel coupling to the $(1,1)$ charge regime, i.e. one electron per dot. Then the graph is initialized by turning on exchange interactions between the desired qubits. This encoding of the qubit-space is key to exploiting fast QND measurement methods measuring in the singlet-triplet basis for measurement-based qubit control of the cluster state. 

\subsection{Exchange Mediated Rotated Measurements} \label{sec:RotMeas}

Consider a simple two DQD spin-qubit example, the first qubit in some arbitrary qubit state $\ket{\psi}=\{\cos(\theta/2),e^{i \phi}\sin(\theta/2)\}$, the second in a singlet state with exchange interaction as described by Eq.~(\ref{eq:Heff}), again assuming identical qubits $=\Delta_{12}=\Delta_{34}=\Delta$. At some time $\tau$ the qubit initially containing $\ket{\psi}$ is measured projectively in the singlet-triplet basis, using the schemes discussed in Sec.~(\ref{sec:CurveMeas}) with outcome $s=0,1$ for a triplet or singlet measurement respectively. The resulting unnormalized state in the $\ket{0}(\ket{1})=\ket{\uparrow\downarrow}(\ket{\downarrow\uparrow})$ basis on the second qubit is 
\begin{equation}
	\ket{\tilde{\psi}}=\frac{1}{\sqrt{2}}\left( 
	\begin{matrix}
		1 & (-1)^s e^{\frac{i  (2 \phi -J\tau)}{2} } \\
		-e^{\frac{i t (2 \phi -J\tau)}{2} } & (-1)^{s+1} e^{2 i \phi}
	\end{matrix}
	\right)\ket{\psi} 
\end{equation}

\noindent where

\begin{equation}
	\phi=\frac{\tau\left(4\mu\Delta+2\mu\Delta_{23}-\sqrt{4\mu\Delta_{23}^2+J_{23}^2}\right)}{2}.
\end{equation}

\noindent Using an exchange $J_{23}=(2n+1)\pi/\tau$ where $n\in \mathbb{Z}$, this simplifies to 
\begin{equation}
	\ket{\tilde{\psi}}=\sigma_x^{s+1} P\left((-1)^{s+1}\theta[\tau]\right) W\left(\theta[\tau]\right)\ket{\psi} 
\label{eq:teleW_withP} \end{equation}

\noindent where 
\begin{equation}
	\begin{split}
			\theta[\tau]=&\frac{\tau}{2} \left(4 \mu\Delta +2 \mu\Delta_{23}-\sqrt{4\mu\Delta_{23}^2+\frac{\pi ^2 (2 n+1)^2}{\tau^2}}\right)\\&+\frac{(2n+1)\pi}{2}.
	\end{split}
\end{equation}

\noindent The resulting projected unitary is the desired exchange controlled single qubit $W(\theta)$ gate, repetitions of which allow for full single qubit control, with additional measurement dependent phase $P\left((-1)^{s+1}\theta\right)$ and Pauli $\sigma_x^{s+1}$ corrections. This additional phase may be corrected after-the-fact by intra-qubit phase gates due to magnetic gradient precession. Importantly however, the $W(\theta)$ control gate is not dependent on the measurement outcome, only the phase correction. The gate time of this projective measurement-based gate depends on the angle $\theta$ of the gate. As a benchmark, to project a Hadamard within this scheme the gate time is $\sim\unit[20]{ns}$\cite{ruskov2019quantum,wardrop2014exchange}, assuming measurement by curvature, discussed in Sec.~\ref{sec:CurveMeas}, that corrects for any leakage from the exchange interaction by projecting only onto the $m_z=0$ subspace, discussed in Sec.~\ref{sec:leak}. It is important to note here that after the application of this protocol, the qubit that initially contained the encoded state is projected onto a state that can be immediately reused in the same protocol. This notion of re-entangling spent qubits of the cluster state helps mediate the overall physical qubit cost of building a processor based on this protocol. This is explored further in Sec.~\ref{sec:Stabiliser}.

%For additional control, let us now consider the same scheme with a control-driving tone $\Omega_d$ applied to the measured qubit applying the phase gate $P(\phi_d)$, along with an additional phase gate $P(\phi_\Delta)$ applied to both qubits by precession due to their magnetic field gradients $\Delta$. The resulting projected unnormalised state is 
%\begin{equation}
%	\begin{split}
%	&\ket{\tilde{\psi}}=\\
%	&\frac{1}{\sqrt{2}}\left( 
%	\begin{matrix}
%		1 & (-1)^s e^{i(2 \tau \mu\Delta+\phi_\Delta+\phi_d)} \\
%		e^{i(2 \tau \mu\Delta+\phi_\Delta)} & (-1)^s e^{-i( \tau J_{23}-4\tau\mu\Delta-2\phi_\Delta-\phi_d)} 
%	\end{matrix}
%	\right)\ket{\psi}.	
%	\end{split}
%\label{eq:phases_matrix} \end{equation}

%\noindent using $\phi_\Delta=2(\pi n_1-\tau\mu\Delta)$ and $J_{23}=\pi(1+2n_2)/\tau$ where $\{n_1,n_2\}\in \mathbb{Z}$ the resulting unitary is simply 
%\begin{equation}
%	\ket{\tilde{\psi}}=\sigma_x^s W\left(\phi_d\right)\ket{\psi}. 
%\label{eq:teleW} \end{equation}
\subsection{Efficient Arbitrary State Preparation} \label{sec:StatePrep}

Here the exchange-measurement hybrid protocols outlined in Sec.~\ref{sec:RotMeas} shall be demonstrated to be useful for single measurement arbitrary state preparation. This is achieved with two qubits, assumed to both start in the spin singlet state ($\ket{-}$ in the qubit basis). The resulting teleported state is after application of (\ref{eq:teleW_withP}) is 
\begin{equation}
	\ket{\psi}= \left( 
	\begin{matrix}
		\sin \left[\frac{\theta}{2}+\frac{\pi}{4}(2+2n+s)\right] \\
		(-1)^{s+1} e^{i \theta} \cos \left[\frac{\theta}{2}+\frac{\pi}{4}(2+2n+s)\right]
	\end{matrix}
	\right) 
\end{equation}

\noindent which can be made to be any arbitrary single qubit state by application of a phase gate.

With this protocol any arbitrary single qubit state can be written with a measurement-based protocol with a single measurement. However, the state is dependent upon the measurement outcome $s$. The two possible states generated by the protocol are always antipodal to one another on the Bloch sphere. To recover the $s=0$ state from a $s=1$ measurement requires a Pauli and phase correction $\sigma_x$ and appropriate phase.

\subsection{Two-qubit Gates} \label{sec:2qub}

\begin{figure}
	[t]
	\centering
	\includegraphics[width=0.9\linewidth]{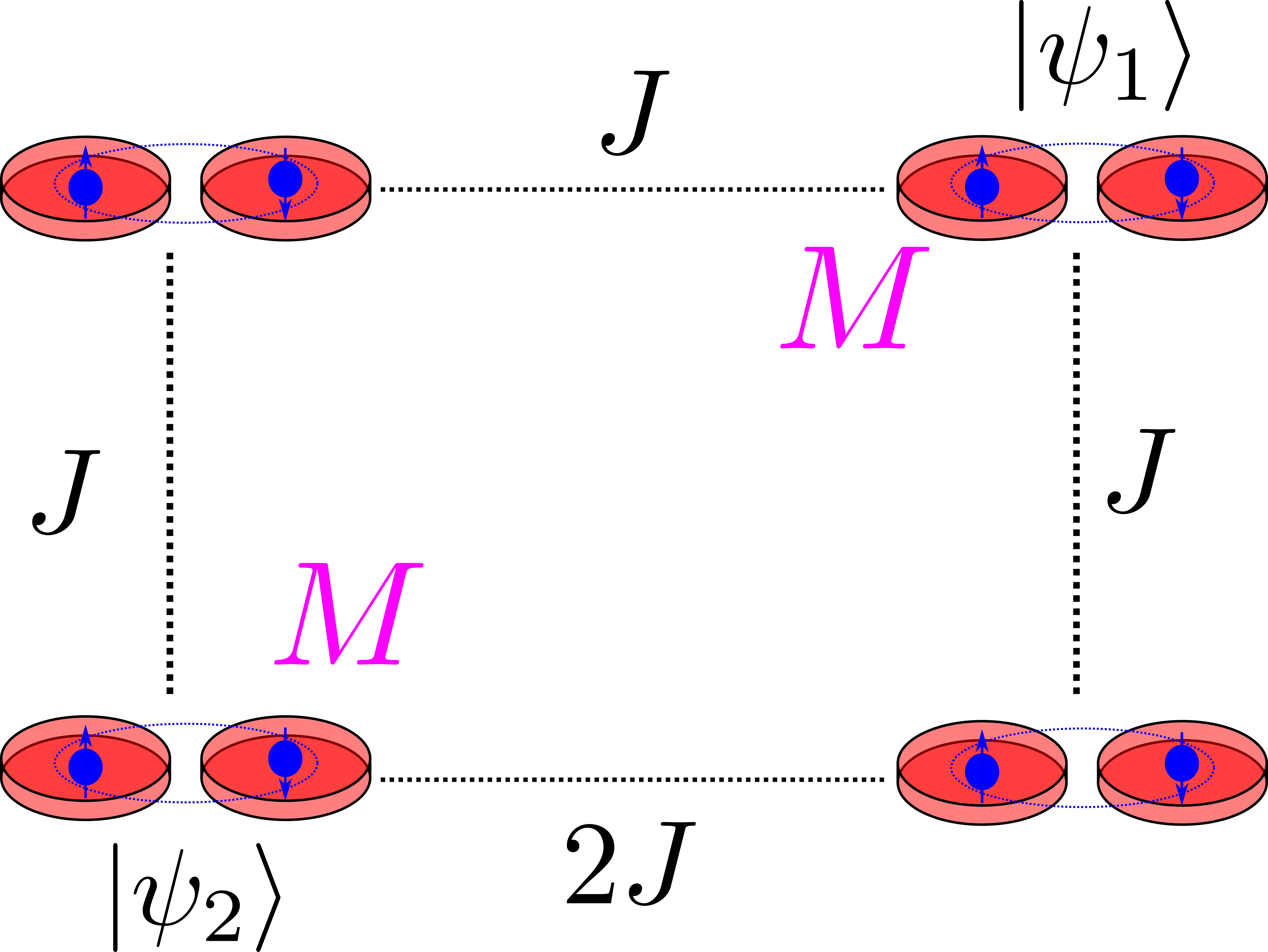}
% \[\Qcircuit @C=1em @R=1em {
%	 \raisebox{-1em}{$(a)$} &&&&&&&&&&&&&&&&&&&&&}\]\\
%	\includegraphics[width=0.7\linewidth]{CXSqu.pdf}
 %   \[\Qcircuit @C=1em @R=1em {
  % 	 \raisebox{-5em}{$(b)$} &&&&&&&&&&&&&&&&&&&&}\]\\
%	 \[\Qcircuit @C=1.5em @R=1.5em {
 %	 &&\lstick{\ket{\psi_1}} & \gate{P_\tau}  & \ctrl{1} & \gate{H} & \gate{Z^{s_1+s_2}} & \qw &\raisebox{-3em}{$\ket{\Psi}$}\\ &&\lstick{\ket{\psi_2}} & \gate{P_\tau} & \targ & \gate{H} & \gate{Z^{s_2}}  & \qw }\]
	\caption{(a) Four-qubit square geometry to perform an entangling two qubit gate in exchange-measurement hybrid cluster state protocol. The initial two qubits start in the top right and bottom left qubits, all qubits are then entangled by exchange to their two neighbors, and singlet-triplet measurements, equivalent to projective Pauli-$x$ measurements, are performed, resulting in an entangling two qubit state in the remaining two qubits. (b) The equivalent gate unitary, including Pauli corrections of the exchange-measurement hybrid protocol given in (a). Here the bottom right qubit in (a) is the first logical qubit and the top left is the second. The measurement outcomes $s_1$ and $s_2$ corresponds to the measurements performed on the physical qubits that initially encoded $\ket{\psi_1}$ and $\ket{\psi_2}$ respectively.} 
	\label{fig:CXSqu}
\end{figure}

In a cluster state, to perform two qubit gates, 2D graphs are necessary. With spin qubits, 2D geometries are limited in scope due to considerations such as plunger-gate placements and cross-talk. However, one of the simplest geometries, four qubits arranged in a square, can be used with the hybrid exchange-measurement graph protocol described here to perform entangling two qubit gates. An example of this is given in Fig.~\ref{fig:CXSqu}. In this protocol, all the qubits are identical, within a global magnetic field gradient from a single micromagnet, such that the intra-qubit field gradients $\Delta_q$ are identical, but the inter-qubit magnetic field gradients $\Delta_{\text{int}}$ depend on the alignment with the two qubits with the global magnetic field gradient. In this example, as in Fig.~\ref{fig:CXSqu}(a), horizontal $\Delta_{\text{int}}$ terms are of equal magnitude and vertical $\Delta_{\text{int}}=0$. Initially the two logical qubits start in the top left and bottom right physical qubits of the square, and the remaining qubits in the $\ket{+}$ state. Inter-qubit interaction is then turned on between all neighboring qubits with time $\tau$ at which point the qubits that previously housed the logical states are measured. The resulting states encoded in the two qubits that weren't measured depend on the values of interaction strength used for each of the four inter-qubit exchange gates. For example, if the value of the interaction strength in Fig.~\ref{fig:CXSqu}(a) used is $J=\pi/\tau$ at times 
\begin{equation}
	\tau=\frac{\pi  (n_2+4 n_1)^2-\pi}{-4 \Delta_{\text{int}} (4 n_1+n_2)}
\end{equation}

\noindent where $n_1 \in \mathbb{Z}, \neq0$ and $n_2=1,5,9\dots$, the resulting two qubit gate performed by this scheme is given in Fig.~\ref{fig:CXSqu}(b). Here it is also assumed that the geometry of the chip is such that the magnetic field gradients satisfy the following

\begin{equation}
	\Delta_q=\frac{\Delta_{\text{int}}(4 n_3 (4 n_1+n_2)+4 n_1+n_2+1)}{2 (4 n_1+n_2-1) (4 n_1+n_2+1)}
\end{equation}

\noindent where $n_3 \in \mathbb{Z}$. The additional phase gates in Fig.~\ref{fig:CXSqu}(b) are $P_\tau=P\left[(\pi/2)\left((4 n_1+n_2)^{(-1)}+2 n_1+n_2\right)\right]$. Note that such a setup of magnetic field gradients and pulse times are only needed to achieve the Clifford circuit in Fig.~\ref{fig:CXSqu}(b), this scheme gives a unitary two-qubit gate so long as three of the exchange gates satisfy $J=\pi/\tau$ and the final gate is $2J$.

\subsection{Re-entangling Measured Qubits} \label{sec:Stabiliser}

\begin{figure}
	[!b] 
	\includegraphics[width=\linewidth]{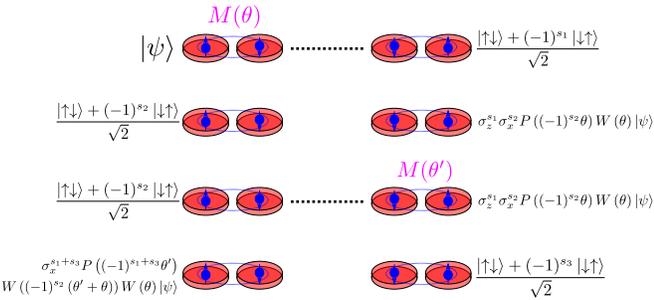} 
	\caption{Example of a simple 2-qubit recycled cluster state with accumulated Pauli corrections. Initially the left hand (LH) qubit is prepared in some state $\ket{\psi}$ and the right hand (RH) qubit is in either a singlet or triplet state. The protocol given in (\ref{eq:teleW_withP}) is then applied to both qubits measuring the LH qubit leaving it in either a singlet or triplet state and the RH qubit in the state given. The protocol is then applied again measuring the RH qubit, adjusting the rotation $\theta'$ of the second application of the protocol based on the phase correction $\theta$ of the previous iteration of the protocol.} 
\label{fig:recycle} \end{figure}

Projective measurements, like that assumed in this work, leave behind a qubit prepared in a singlet or triplet state which may be re-entangled to the resource state. By again engaging the exchange interaction $J_{23}$ the qubit may be re-entangled with the information carrying qubit to allow for repetition of a measurement-based gate on the same geometry. This circumvents the issues of resource overhead for solid state implementations of MBQC for non-commuting measurements. The adjustments needed to do so depend upon the previous measurement result. As such the projected gate given by the protocol~(\ref{eq:teleW_withP}) becomes 
\begin{equation}
	\ket{\tilde{\psi}}=\sigma_z^{s_1+1} \sigma_x^{s_2} P\left((-1)^{s_2}\theta\right) W\left(\theta\right)\ket{\psi} 
\label{eq:teleW_withP_s2} \end{equation}

\noindent where $s_1$ is the result of the previous measurement performed on the qubit the state is being projected to and $s_2$ is the result of the current measurement projecting the state back onto the first qubit. Performing multiple gates with this protocol, the overall number of post gate corrections, i.e. up to two Pauli corrections and one phase correction, remains constant regardless of the number of iterations if the previous known anomalous phase is considered when performing the next gate. For example, assuming an initial two qubit state $\ket{\psi}\otimes\ket{(-1)^{s_1}-}$, if the protocol~(\ref{eq:teleW_withP}) is repeated twice for two different angles $\theta$ and $\theta'$ respectively, the applied unitary is 
\begin{equation}
	U=\sigma_x^{s_1+s_3} P\left((-1)^{s_1+s_3}\theta'\right)W\left((-1)^{s_2}\left(\theta'+\theta\right)\right) W\left(\theta\right). 
\label{eq:teleW_withP_multiple} \end{equation}

\noindent This shows that subsequent rotations $\theta'$ may be adjusted to account for the measurement outcomes of the previous iteration of the protocol to apply the desired series of unitaries without performing a phase correction at each step. Note that if the protocol is applied an odd number of times, such as in~(\ref{eq:teleW_withP_multiple}), a maximum of one Pauli-$x$ correction may be needed, whilst if the protocol is applied an even number of times, a maximum of Pauli-$x$ and an Pauli-$z$ correction may be needed. 
%For the protocol given in~(\ref{eq:teleW}), the effect of the additional Pauli-$z$ correction as in~(\ref{eq:teleW_withP_s2}) with re-entangled qubits is also observed.

\begin{figure}
	[!t] 
	\includegraphics[width=\linewidth]{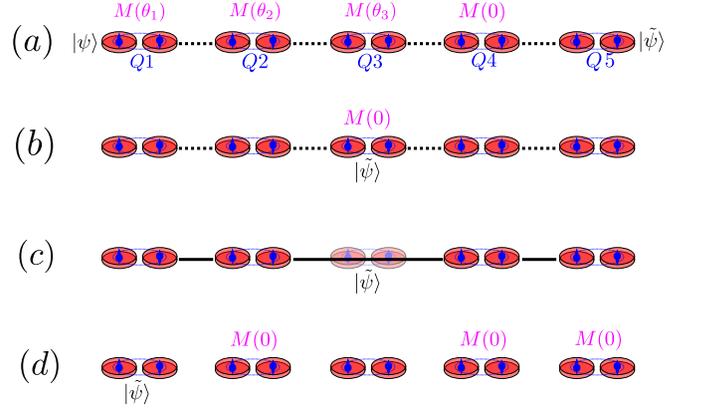} 
	\caption{(a) Linear cluster state geometry with measurements (magenta) to perform an arbitrary qubit rotation $U$ on state $\ket{\psi}$ initially encoded in qubit $Q1$. The final state $\ket{\tilde{\psi}}=U\ket{\psi}$ is teleported to qubit $Q5$. The black dashed lines denote exchange interaction to allow for necessary measurements. (b) Graph state geometry to encode the state $\ket{\tilde{\psi}}$ to the $[[4,1,2]]$ stabilizer state given by the solid black lines in (c), by measuring $Q3$. (d) Protocol for decoding the stabilizer state given in (c), projected $\ket{\tilde{\psi}}$ to $Q1$.} 
\label{fig:rot_stab_decode} \end{figure}

With an architecture in which each qubit can be measured and re-entangled to the graph on demand, a processor which can be used both for qubit operations and quantum error correction encoding can be designed. Graph states are naturally compatible with stabilizer codes as the stabilizers $K_i$ of a cluster state may be inferred from its geometry as 
\begin{equation}
	K_i=(\sigma_x)_i \otimes_{j\in N(i)}(\sigma_z)_j. 
\end{equation}

\begin{figure}
	[!t] 
	\includegraphics[width=0.75\linewidth]{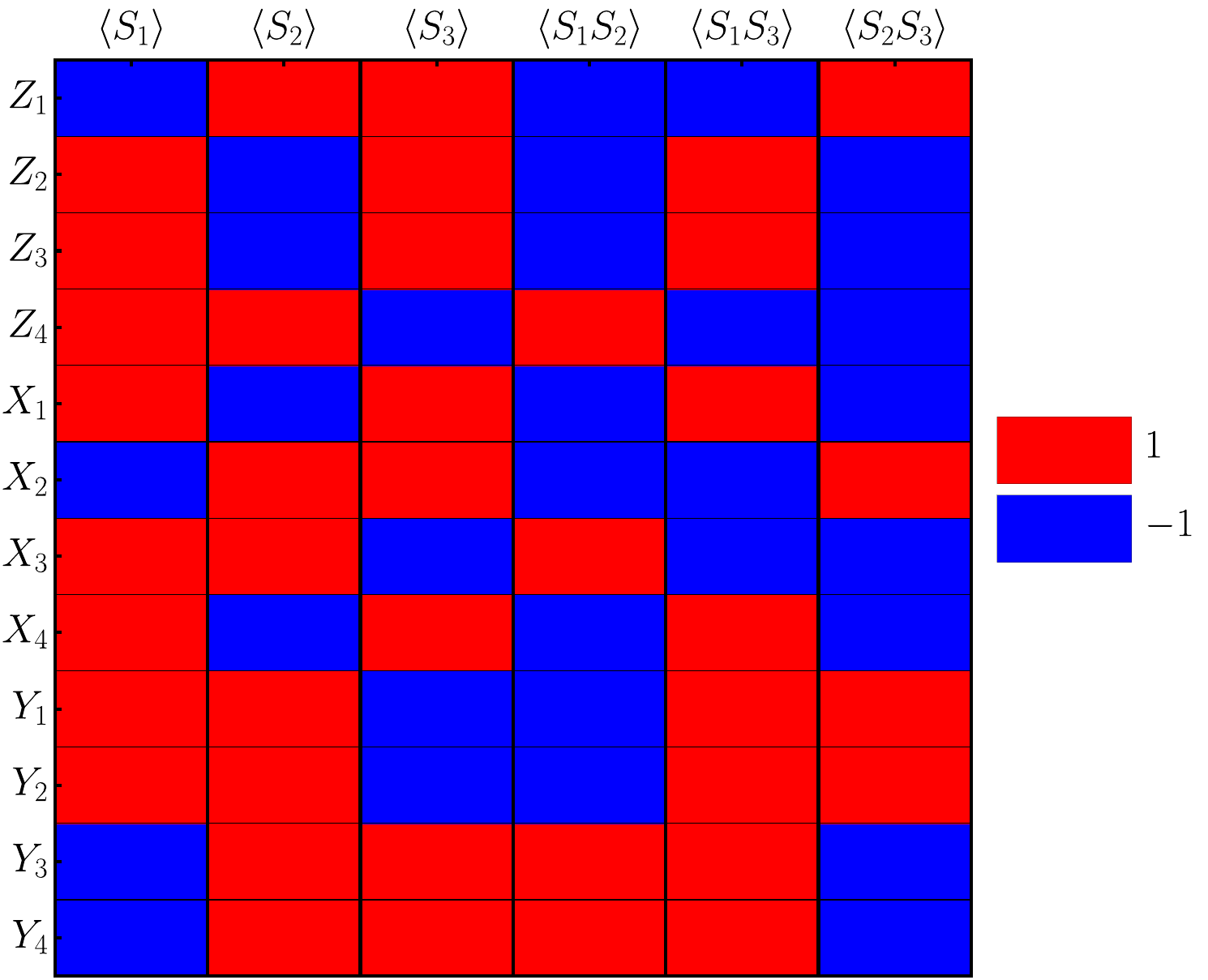} 
	\caption{Table of the expectation values of the stabilizers $\{S_1,S_2,S_3\}$ and products of the linear stabilizer code given in Fig.~\ref{fig:rot_stab_decode} under every single qubit Pauli on the $i$\textsuperscript{th} error $\{X_i,Y_i,Z_i\}$. The occurrence of any single qubit error is detectable, but some are indistinguishable from a single measurement of the syndromes.} 
\label{fig:StabTab} \end{figure}

\noindent For example, encoding and decoding a qubit with a $[[4,1,2]]$ stabilizer state has been demonstrated experimentally with a five qubit cluster state in an optical system\cite{bellExperimentalDemonstrationGraph2014}. The graph geometry used for this code is not feasibly compatible with a spin-qubit implementation. However, a similar, linear code can be used, at the cost of distinguishing some single qubit Pauli errors. This is achieved with an architecture of five qubits in a line. The same architecture is described in Sec.~\ref{sec:MBQC} to achieve arbitrary single qubit rotations of the form 
\begin{equation}
	\ket{\tilde{\psi}}=W(0)W(\theta_3)W(\theta_2)W(\theta_1)\ket{\psi} 
\end{equation}

\noindent with the rotated state ending in the fifth physical qubit. This state may then be transferred back to the middle (third) qubit by two $W(0)$ measurements. Then, by entangling all adjacent qubits with a $CZ$ interaction and measuring the third qubit in the Pauli-$x$ basis, $\ket{\tilde{\psi}}$ is then encoded in a $[[4,1,2]]$ stabilizer state across the remaining four qubits. The qubit can then be recovered to the first qubit form the stabilizer to perform additional rotations by measuring the second, fourth and fifth qubit in the Pauli-$x$. This protocol, depicted in Fig.~\ref{fig:rot_stab_decode}, has stabilizers $\{S_1,S_2,S_3\}=\{\sigma_x\otimes\sigma_z\otimes\mathbb{I}\otimes\mathbb{I},\sigma_z\otimes\sigma_x\otimes\sigma_x\otimes\sigma_z,\mathbb{I}\otimes\mathbb{I}\otimes\sigma_z\otimes\sigma_x\}$, and can detect any arbitrary single qubit Pauli error, though cannot distinguish between all such errors, see Fig.~\ref{fig:StabTab}. 

%The only missing ingredient in the solid state proposal given here is a non-destructive measurement in the Pauli-$z$ basis. This can be achieved destructively with spin-to-charge conversion. Alternatively, a QND Pauli-$z$ measurement can be achieved with an ancilla qubit and the same QND measurement assumed for the rest of this work. This is achieved by entangling the target qubit by a $CZ$ operation to an ancilla in a triplet state, i.e. $\ket{+}$ in the qubit basis, as is assumed in Sec.~\ref{sec:RotMeas}, then measuring the ancilla qubit in Pauli-$x$ basis.

\section{Entangling Measurements Protocol} \label{sec:Quantum Bus}

The methods presented up to this point are a form of MBQC whereby the measurements provide unitary operations on the encoded qubits, at the cost of entanglement between qubits generated by exchange gates. This allows for a powerful method of processing and encoding quantum information with an at worse polynomial cost in physical qubits, which can be mediated by re-entangling measured qubits. The following protocol shown here, however, implements measurements that entangle qubits, as well as measurements that rotate qubits at the cost of entanglement. 

\subsection{Qubit Model: No Magnetic Gradients} \label{sec:nogradqub}

In this scheme qubits are encoded in the Pauli-$z$ basis of the singlet-triplet spin space ($\ket{0}=\left(\ket{\uparrow\downarrow}-\ket{\downarrow\uparrow}\right)/\sqrt{2}$ and $\ket{1}=\left(\ket{\uparrow\downarrow}+\ket{\downarrow\uparrow}\right)/\sqrt{2}$) and require only controllable exchange coupling between dots within the same qubit. No magnetic field gradients or inter-qubit exchange coupling are required. Therefore, the effective Hamiltonian for the inter- and intra-qubit exchange interaction of two qubits comprised of four dots (as in Fig.~\ref{fig:DQD}(b)) can be written as\cite{li2012controllable}

\begin{equation}
		H_{\text{eff}}=\frac{J_{12}}{2}\sigma_z^1+\frac{J_{34}}{2}\sigma_z^2-\frac{J_{23}}{4}\sigma_x^1\sigma_x^2.
		\label{eq:Heff_nograb}
\end{equation} 

\noindent Note that since there are no magnetic field gradients considered here, if the inter-qubit exchange term $J_{23}>0$ leakage to the $m_z=\pm1$ triplet states will occur. Although this leakage will be suppressed by qubit measurement, it will be assumed that $J_{23}=0$, as this term does not contribute to the entanglement generating steps in this protocol, the inter-qubit measurements.

\subsection{Inter-Qubit Measurement}\label{sec:interqubit}

Alternatively to measuring in the basis of a logical qubit, an entangling operation can be achieved by measuring a dot from each adjacent encoded qubit, i.e. measuring in the singlet-triplet basis of the second and third QDs in Fig.~\ref{fig:DQD}. Forgoing any exchange interaction, the resulting 2-qubit projector in the qubit basis used in Sec.~\ref{sec:Hybrid} is 
\begin{equation}
	\mathcal{M}=\ket{00}\bra{00}+\ket{11}\bra{11}.
\label{eq:2qubitop} \end{equation}
\noindent However, this measurement also entangles the encoded qubits to spin-states outside of the $m_z=\pm1$ triplet states. The measurement cannot however change the net-spin state of the two qubits it acts on and, as such, measuring either of the individual qubits in the $m_z=0$ singlet-triplet basis corrects for the leakage, at the cost of the entanglement of the qubits.

%A simple example of a use of the projector in Eq.~(\ref{eq:2qubitop}), in this qubit basis, is as a quantum bus. This is achieved by encoding qubit one in some arbitrary state $\ket{\psi}$ and qubit two in a $\ket{+}$ or singlet state. By first performing the entangling measurement between the two qubits, then measuring qubit one, the resulting state given to the second qubit is $\sigma_z^s\ket{\psi}$ where $s$ is the result of the second measurement and the leakage is corrected for. With the exchange interaction and phase gates considered in Sec.~\ref{sec:Hybrid}, the projected state after both measurements includes a phase gate $\ket{\tilde{\psi}}=\sigma_z^sP\left(-J_{23} \tau+4 \mu\Delta t+2 \phi_\Delta\right)\ket{\psi}$. In this basis however, full qubit control cannot be performed with measurements of this type. 

\subsection{Qubit Control with Entangling Measurements}\label{sec:fullEntMeasCont}

The entangling measurement described can be used along with intra-qubit exchange-interaction to implement an alternative non-cluster state approach to MBQC. In this regime, as shown in ~(\ref{eq:Heff_nograb}), again assuming the inter qubit exchange $J_{23}=0$, the intra-qubit exchange terms $J_{12(34)}$ allow for phase control in the qubit subspace. The final tool needed for single and two qubit control is access to an ancilla qubit initialized in the $\ket{+}=\ket{\uparrow\downarrow}$ state for each logical qubit (polynomial increase in physical qubits) and the ability to measure each qubit in the singlet-triplet basis.

%\begin{equation}
%	P(J,\tau)=\left( 
%	\begin{matrix}
%		1 & 0 \\
%		0 & e^{i J \tau}
%	\end{matrix}\right)
%	\label{eq:Rz}
%\end{equation}

\begin{figure}
	[!b]
	\centering
	\includegraphics[width=\linewidth]{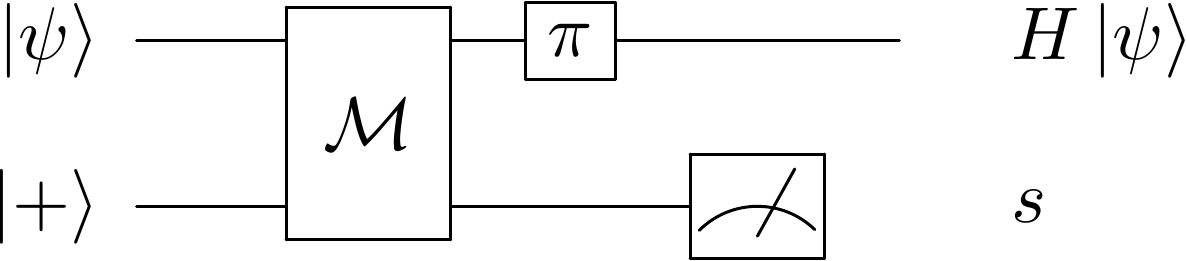} 
	\caption{Circuit diagram of the phase gates and measurement sequence needed to achieve a single qubit Hadamard without any correction.} 
	\label{fig:single_qubit_noB}
\end{figure}

With these tools, a complete set of single qubit unitaries can be produced as follows. Suppose initially there are two qubits, the first initialized to some arbitrary state $\ket{\psi}$ and the second in the $\ket{+}$ state. Then both qubits experience an exchange phase gate imparting phases $\phi_1$ and $\phi_2$ to the respective qubits, before which an entangling measurement is performed on the qubits and after which another set of phase gates $\phi_3$ and $\phi_4$ are performed before the second qubit is measured with outcome $s$. This results in the following unitary acting on the input state in the first qubit

\begin{equation}
	\ket{\tilde{\psi}}=\frac{1}{\sqrt{2}}\left( 
	\begin{matrix}
		1 & e^{i (\phi_1 + (-1)^s \phi_2)} \\
		e^{i ((-1)^s\phi_1 + \phi_2)} & e^{i (\phi_2 + \phi_3)} 
	\end{matrix}\right)\ket{\psi}.
\end{equation}

\begin{figure}
	[!t]
	\centering
	\includegraphics[width=\linewidth]{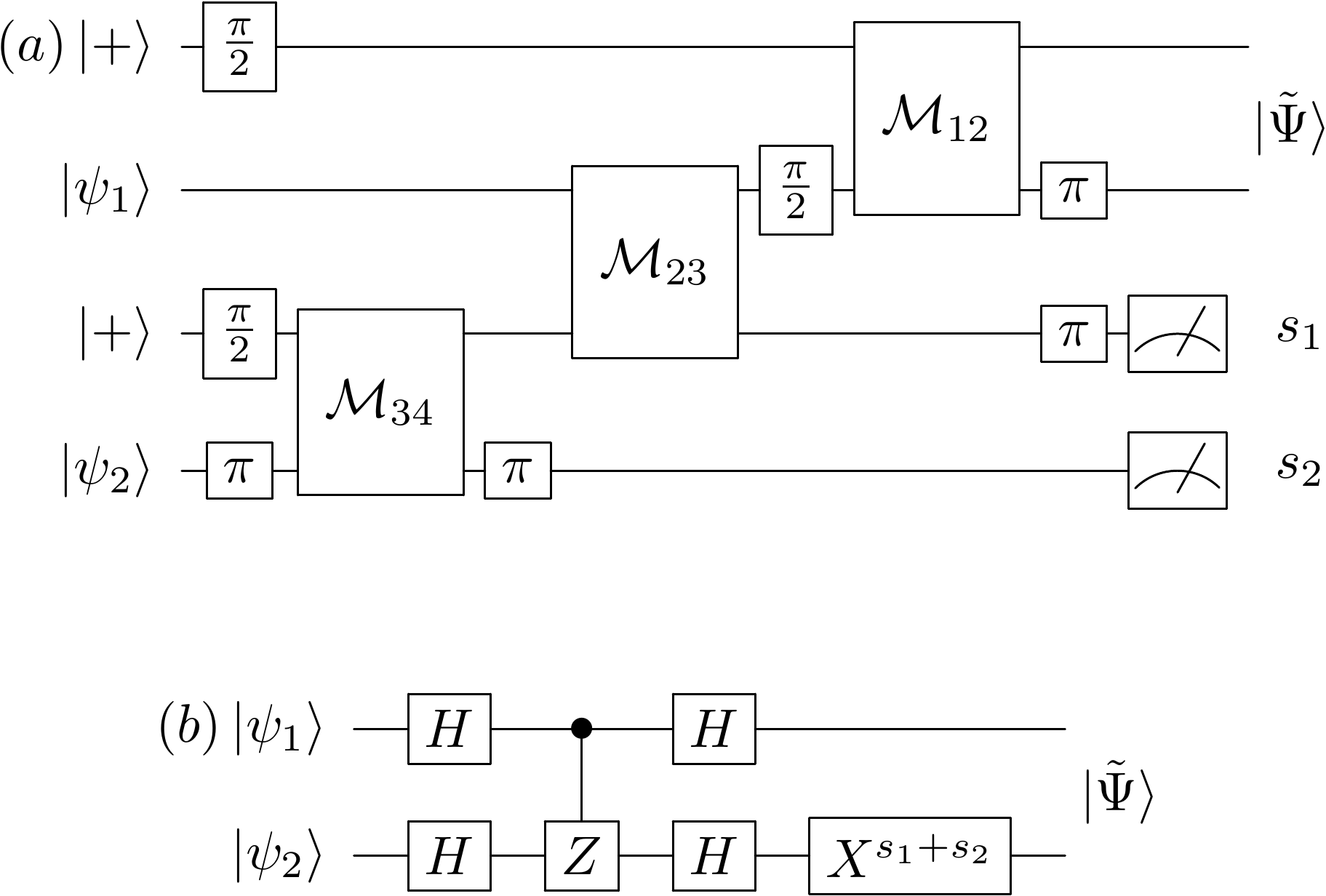} 
	\caption{(a) Circuit diagram of the phase gates ($\pi$ or $\pi/2$) and measurement sequence needed to achieve a two qubit entangling unitary with two ancilla qubits. (b) Circuit diagram of the two qubit unitary performed by the measurement and phase sequence given in (a) with single Pauli-$x$ correction.} 
	\label{fig:two_qubit_noB}
\end{figure}

\noindent Then by selecting appropriate phases, useful gates up to some known correction may be implemented. For example, Fig.~\ref{fig:single_qubit_noB} shows the sequence of phase gates and measurements needed to perform a Hadamard gate on the logical qubit $\ket{\psi}$ without any Pauli correction. The only measurement outcomes that determine the correction is the ancilla qubit measurement $s$ at the end of the sequence; the entangling measurement result does not contribute. The same tools can be used to implement a two qubit entangling gate with two ancilla qubits. The measurement and phase gate sequence needed is given in Fig.~\ref{fig:two_qubit_noB}(a). The resulting two qubit unitary of the sequence given in Fig.~\ref{fig:two_qubit_noB}(a) is shown in Fig.~\ref{fig:two_qubit_noB}(b) including the single Pauli-$x$ correction given by the measurement results $s_1+s_2$ of the measurement of the ancilla. This is a demonstration of implementing a Clifford gate group with the proposed scheme. This is done for simplicity to demonstrate that a universal gate set can be implemented, however, non-Clifford two qubit gates are possible with the same tools.

For the single qubit gate scheme the entangling measurement with the measurement on the ancilla qubit does not allow for leakage of the logical qubit, regardless of the phase gates used and the or ancilla qubit initial state. The two qubit gate example given also surpreses leakage, however, this is due to the selection of the phase gates, initial ancilla states, and measurement sequence and is therefore a significant factor to consider when designing gates within this scheme. The gate times of both the single and two qubit gates are limited by the single qubit phase gates. Assuming an exchange coupling of $J=\unit[160]{MHz}$\cite{reed2016reduced} gate times of $\sim\unit[80]{ns}$ for the Hadamard and $\sim\unit[140]{ns}$ for the rotated $CZ$ gates shown in Fig.~\ref{fig:single_qubit_noB} and Fig.~\ref{fig:two_qubit_noB} respectively.

This scheme offers a method of hybrid exchange measurement-based approach to quantum computation similar to a traditional gate-based approach without the need for magnetic field gradients or driving fields and can be made to be robust to leakage as well as operated at the SOP of the physical qubits. The main limitation is the need for ancilla qubits initialized to the $\ket{+}$ state as this method offers no way of preparing such states from a $\ket{0}$ or $\ket{1}$ state. Such states can be initialized by adiabatically turning off the exchange interaction\cite{foletti2009universal,fogarty2018integrated} of a qubit in the triplet state, which is within the control parameters of the scheme. Unfortunately however, since this scheme is more akin to traditional gate-based approaches to quantum computations, the inherent capability of the scheme with stabilizer codes described in Sec.~\ref{sec:Stabiliser} does not apply to this scheme.

\section{Measurement Methods} \label{sec:CurveMeas}

In this work two schemes for MBQC with singlet-triplet spin qubits are given assuming only measurement of the singlet-triplet basis. This can be achieved in a number of different ways, however here we shall consider two such methods. The conventional spin-to-charge method and the QND curvature method.

The conventional spin-to-charge method of measuring in the singlet-triplet in achieved by adiabatically tuning the tunnel coupling between the two dots being measured. Due to Pauli-exclusion, the singlet state allows for hopping from the $(1,1)$ to the $(0,2)$ while the triplet state does not couple to the doubly occupied state\cite{hanson2007spins}. If the singlet state is measured, it can then be re-initialized back to the $(1,1)$ singlet state by again adiabatically tuning the tunnel coupling.

Alternatively, the singlet-triplet basis can be measured by coupling to a superconducting (SC) resonator\cite{ruskov2019quantum}. The resonator is driven with frequencies $\omega_m\approx\omega_r$ and $\omega_s\approx\omega_q$ with strength $\epsilon_s$, where $\omega_r$ is the frequency of the resonator, $\omega_m$ is the frequency of an external gate modulation near the resonator and $\omega_q$ is the singlet-triplet qubit frequency. This scheme exploits the quantum curvature of the qubits and is described as follows in the rotating frame and qubit-space of the qubit-qubit interaction Hamiltonian~(\ref{eq:Heff}) 
\begin{equation}
	\begin{split}
		H_{\text{meas}}=(\omega_r-\omega_m)a^\dagger a+\frac{(\omega_q-\omega_s)}{2}\sigma_z+&\\(\chi+\delta\omega) a^\dagger a\sigma_z+\frac{1}{2}(g_\parallel\sigma_z+g_0\sigma_0)(a+a^\dagger)&\\+\epsilon_d(a e^{-i\omega_m t}+a^\dagger e^{i\omega_m t})& 
	\end{split}
	\label{eq:Measure} \end{equation}

\noindent where $a(a^\dagger)$ are the annihilation (creation) operators of the resonator, $g_\parallel$ is the longitudinal coupling between resonator and qubit and $g_0$ the static coupling. Here the important term for measuring the state of the qubit is the $\Delta\omega$ term. This term is the change in the SC resonator frequency due to the state of the qubit. This term is given by the quantum curvature of the eigen-energies of the qubit and is defined as 
\begin{equation}
	\delta\omega=\hbar\omega_r^2\left(\frac{\eta}{e_0}\right)^2\frac{\partial^2 G(V_m^0)}{\partial V_m^{2} } 
\end{equation}

\noindent where $\omega_r$ is the resonator frequency, $\eta$ is the dimensionless parameter for resonator inductance and $G(V_m^0)$ is defined as follows, assuming the singlet-triplet qubit Hamiltonian as 
\begin{equation}
	\begin{split}
		H&=E_s\ket{s}\bra{s}+E_{t_0}\ket{t_0}\bra{t_0}\\
		&=G(V_m^0)\mathbb{I}+C(V_m^0)\sigma_z. 
	\end{split}
\end{equation}

Measuring $\delta\omega$ is assumed to quickly perform a projective measurement in the singlet-triplet spin basis when the cavity is driven.

\section{Leakage Protection} \label{sec:leak}

Supposing input states free of leakage, any leakage during the protocols, either by exchange or inter-qubit measurement, will be corrected for by projective single qubit measurements in the $m_z=0$ subspace of all ancillae qubits, in both protocols, provided negligible scattering of the total $m_z$ of the all the entangled qubits. The only channel for leakage of the final state of data qubit(s) after all the gate protocols considered here is measurement leakage.

Both methods of measurement can be used for the two MBQC protocols given here. One major difference between the two methods is inherent protection against leakage. In the curvature measurement scheme, the $m_z=\pm1$ states are energetically separated from the cavity frequency by an external magnetic field, however in the spin-to-charge method, measurement of a triplet state is not exclusive to the $t_0$ state, but a projection onto all triplet states. This allows for measurement induced leakage. Fortunately for MBQC, the outcome of the measurements of these is only consequential to the Pauli-correction terms. Consequently, so long as valley excitations are avoided, the measurement need may instead be replaced with initializing the qubit into the $(0,2)$ singlet state by decay or by varying the qubit detuning. For MBQC to function, the measured ancillae need to be completely removed from the entangled system, which in the case of singlet-triplet qubits means the two spins of the measured qubit be maximally entangled. Since projection onto a combination of all triplet states considered here does not completely remove the entangled qubit from the system, the measurement is in a sense incomplete. However, since the measurement is incomplete, the leakage on the system from is addressable. For example, in the first protocol, for a single qubit rotation by measurement given in (\ref{eq:teleW_withP}), suppose the measurement of the ancilla is done by spin-to-charge and the result is a triplet. After this measurement, the gate has not succeeded, but can be recovered by setting the detuning of the ancilla qubit back to the operating point and allowing the qubits to precess for time

\begin{equation}
	\tau=\frac{4(\pi + 4\pi n)}{\sqrt{\tilde{J_{23}}^2+4\Delta_{23}^2}-4\Delta_{23}}
\end{equation}

\noindent where $n\in \mathbb{Z}$, then measuring the qubit again. Note that during this step it assumed that the inter-qubit exchange $J_{23}=0$, the term $\tilde{J_{23}}$ is the exchange used in the previous pulse before the failed triplet measurement. The result of the following measurement must be a singlet, i.e. the measurement will be successful, and the only difference between the resulting rotation and the intended rotation is an additional Pauli-$z$ correction term.  

\section{Discussion} \label{sec:Discussion}

Here protocols for full measurement-based qubit control with Si DQD qubits using two hybrid measurement-exchange schemes are given. The first protocol is given by entangling neighboring qubits with an exchange interaction and single qubit measurements resulting in spin-qubit cluster-state approach to MBQC. This allows for full single and two-quit control with known Pauli and phase corrections. The leakage limitations of inter-qubit exchange gates do not apply to this protocol given projective single qubit measurements without leakage, or by the correction steps given in Sec.~\ref{sec:CurveMeas}. This scheme is shown to also be useful for efficient arbitrary qubit state preparation. The nature of the measurements allows for the limiting of the required qubit resource overhead by recycling measured qubits back into the resource state with additional known required Pauli corrections. This allows for simple processor architectures designed for the arbitrary manipulation as well as encoding and decoding of a qubit state into stabilizer state with five physical qubits in a 1D geometry.

A second protocol with entangling measurements is also shown by measuring in the singlet-triplet basis between one dot from one qubit with one dot from a different qubit. This measurement-based protocol is closer to a traditional gate-based protocol, and can be made to be robust against leakage and can be operated at the SOP of all the physical qubits, requiring only measurements and phase gates. Again, although the entangling measurements couple to states outside of the qubit subspace, single qubit measurements limited to qubit subspace will remove all leakage from the end result of each gate.

\begin{table}
	[t]
\begin{center}
\begin{tabular}{ l|c|c|c|c } 
   & Protocol 1 & $S/T_0$ & Protocol 2 & 3-Spin \\
 \hline
 \hline
 \multicolumn{5}{c}{Single Qubit Gate} \\
 \hline
 Gates & 1 & 1 & 3 & 4 \\ 
 Measurements & 1 & - & 2 & - \\ 
 Ancillae & 1 & 0 & 1 & 0 \\
 Leakage Protected & Yes & No & Yes & No \\
 \hline
 Correction Gates & - & 7 & - & 30 \\
 Correction Ancillae & - & 1 & - & 1 \\ 
 \hline
 \multicolumn{5}{c}{Two Qubit Gate} \\
 \hline
 Gates & 4 & 1 & 7 & 22 \\ 
 Measurements & 2 & - & 5 & - \\
 Ancillae & 2 & 0 & 2 & 0 \\ 
 Leakage Protected & Yes & No & Yes & Yes \\
 \hline
 Correction Gates & - & 14 & - & - \\
 Correction Ancillae & - & 2 & - & - \\
\end{tabular}
\end{center}
\caption{Table comparing the number of exchange gates, timesteps, measurements and ancillae between the first proposed cluster state protocol, and leakage protected operations in singlet-triplet qubits with magnetic field gradients from literature\cite{mehl2015fault} and the second entangling measurements protocol and gradient free exchange only three spin-encoded qubits from literature\cite{fong2011universal}. Single qubits operations from literature aren't robust to leakage, so given numbers include leakage correction circuits after a rotation. There are no two qubit leakage robust sequences in the literature, only leakage suppressed, so for comparison the leakage correction circuit is considered twice. Both proposed protocols offer significant improvements over their similar respective counterparts.} 
\label{tab:comp_pulse}
\end{table}

In the first protocol, a single single qubit gate consists of one exchange gate and one measurement, whilst a two qubit control consists of four exchange gates and two measurements, all performed simultaneously, assuming measurement by curvature. In the second protocol single qubit gates consist of three exchange gates and two measurements whilst two qubit gates consist of seven exchange gates, five measurements and $n$ ancilla qubits per $n$ data qubits (total $2n$ physical qubits), again assuming measurement by curvature. The first proposed protocol is an improvement compared to conventional gate-based leakage resistant schemes in singlet-triplet encoded qubits with a magnetic field gradients, where single qubit leakage errors can be corrected with a minimum of five exchange gates and one ancilla qubit\cite{mehl2015fault}, and two qubit gates are assumed to be given by long range cavity coupling, unlike in the work presented. Additionally, the second proposed protocol is more efficient than leakage protection and correction in similar systems without magnetic field gradients, three-spin exchange-only qubits. In such systems, active leakage reduction requires 30 exchange gates and protected two qubit gates require a minimum of 22 exchange gates\cite{fong2011universal}, which can be improves by chip geometry or with dynamic gates\cite{setiawan2014robust,hickman2013dynamically}. A direct comparison of these schemes is made in Table~\ref{tab:comp_pulse}. Equally, both schemes are superior when compared to superconducting systems that can also experience leakage errors, requiring 16 single and two qubits gates and two measurements\cite{ghosh2015leakage}.

This work promises alternative hybrid exchange-measurement-based quantum computation schemes. The estimated gate times for all the schemes proposed here are $<\unit[150]{ns}$ and so are all well within expected relaxation times of Si DQD spin qubits of tens of microseconds\cite{sigillito2019site}. This allows for semiconductor qubits to take advantage of MBQC without the resource overhead or necessity for arbitrary rotated measurement schemes, and with the addition of control by exchange, allows for more deterministic control compared to other proposals\cite{freedman2021symmetry}.

\section{Acknowledgements} \label{ref:Acknowledgements}

We acknowledge helpful discussions with M. Benito, R. Ruskov, M. Russ, YP. Shim, V. Shkolnikov and Y. Yanay.

\end{document}